\documentclass[11pt]{article}
\usepackage{coling2020}
\usepackage{times}
\usepackage{url}
\usepackage{latexsym}
\usepackage{booktabs}
\usepackage{graphicx}
\usepackage{amsmath, amsfonts}
\usepackage{amsthm}
\usepackage{multirow}
\usepackage{tikz}
\usetikzlibrary{shapes.geometric, arrows}

\newtheorem{theorem}{Definition}
\colingfinalcopy 

\title{Dual Attention Model for Citation Recommendation}

\author{Yang Zhang \\
  Graduate School of Informatics \\ Kyoto University \\
  {\tt zhang.yang.33z@st.kyoto-u.ac.jp} \\\And
  Qiang Ma \\
  Graduate School of Informatics \\ Kyoto University \\
  {\tt qiang@i.kyoto-u.ac.jp} \\}

\date{}

\begin{document}
\maketitle
\begin{abstract}
\blfootnote{
    %
    %
    %
    %
    %
    %
    \hspace{-0.65cm}  
    This work is licensed under a Creative Commons 
    Attribution 4.0 International License.
    License details:
    \url{http://creativecommons.org/licenses/by/4.0/}.
}
Based on an exponentially increasing number of academic articles, discovering and citing comprehensive and appropriate resources has become a non-trivial task. Conventional citation recommender methods suffer from severe information loss. For example, they do not consider the section of the paper that the user
is writing and for which they need to find a citation, the relatedness between the words in the local context (the text span that describes a citation), or the importance on each word from the local context. These shortcomings make such methods insufficient for recommending adequate citations to academic manuscripts. In this study, we propose a novel embedding-based neural network called ``dual attention model for citation recommendation (DACR)'' to recommend citations during manuscript preparation. Our method adapts embedding of three semantic information: words in the local context, structural contexts \footnote{Cited papers other than the target citation in a citing paper, which are defined in \cite{9123859} and Definition \ref{def2} in Section \ref{sec:definition} of this paper.}, and the section on which a user is working. A neural network model is designed to maximize the similarity between the embedding of the three input (local context words, section and structural contexts) and the target citation appearing in the context. The core of the neural network model is composed of self-attention and additive attention, where the former aims to capture the relatedness between the contextual words and structural context, and the latter aims to learn the importance of them. The experiments on real-world datasets demonstrate the effectiveness of the proposed approach.
\end{abstract}

\section{Introduction}
\label{intro}
%
%

\begin{figure}[tb]
  \centering
  \includegraphics[width=0.7\textwidth]{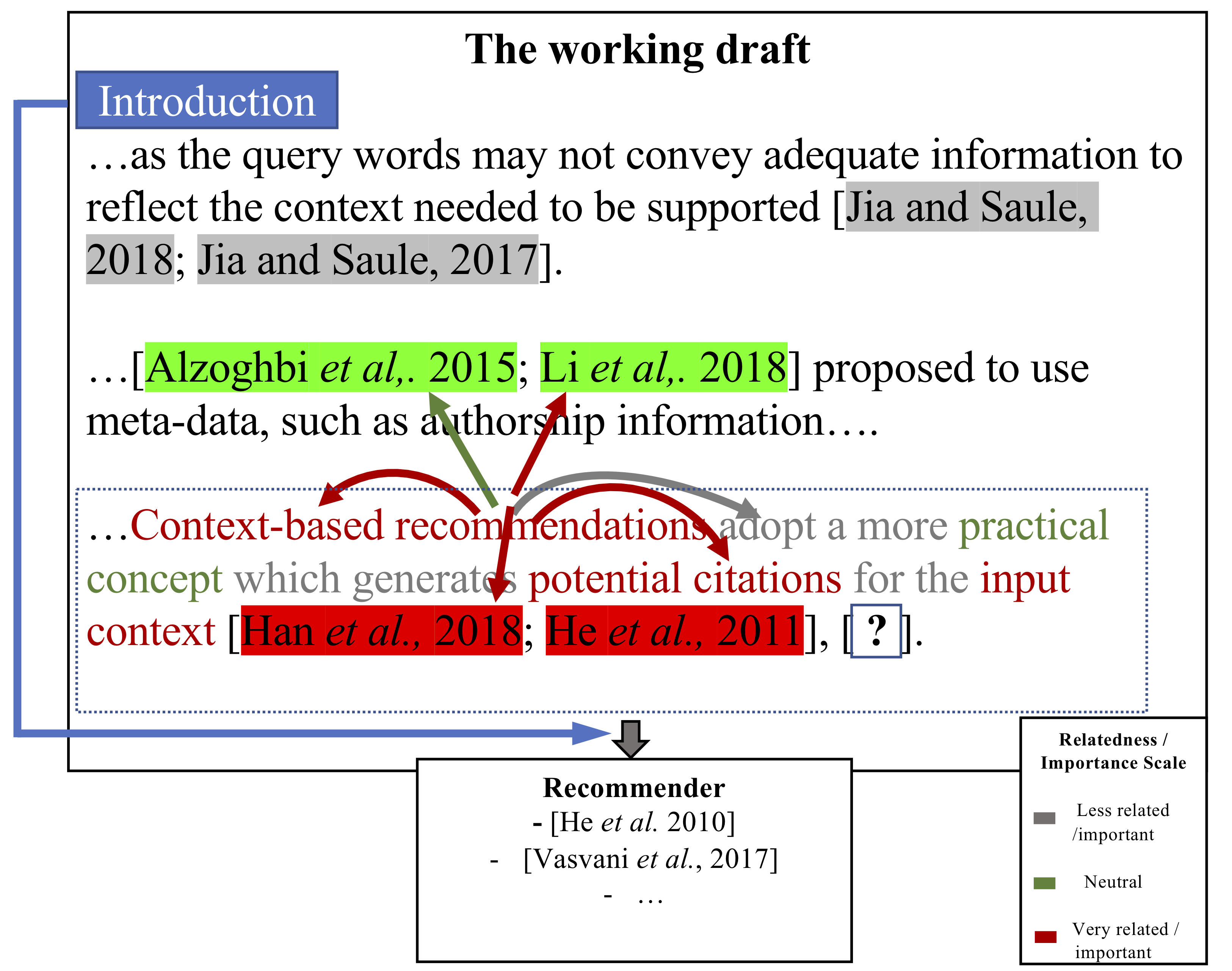}
  \caption{Concept of dual attention model for citation recommendation (DACR)}
  \label{fig:concept}
\end{figure}

When writing an academic paper, one of the most frequent questions considered is: ``Which paper should I cite at this place?'' Based on the massive number of papers being published, it is impossible for a researcher to read every article that might be relevant to their study. Thus, recommending a handful of useful citations based on the contents of a working draft can significantly alleviate the burden of writing a paper. An example of the application scenario is demonstrated in Figure \ref{fig:concept}. 

Currently, many scholars rely on ``keyword searches'' on search engines, such as Google Scholar \footnote{https://scholar.google.com/} and DBLP \footnote{https://dblp.uni-trier.de/}. However, keyword-based systems often generate unsatisfying results, because query words may not convey adequate information to reflect the context that needs to be supported \cite{DBLP:conf/asunam/JiaS17,DBLP:conf/ecir/JiaS18}. Researchers in various fields have proposed various methods to solve this problem. For example, studies in \cite{DBLP:conf/cscw/McNeeACGLRKR02,DBLP:conf/webi/GoriP06,DBLP:conf/jcdl/CarageaSMG13,DBLP:conf/asunam/KucuktuncSKC13,DBLP:conf/ecir/JiaS18} considered recommendations based on a collection of seed papers, and \cite{DBLP:conf/lwa/AlzoghbiA0L15,DBLP:journals/access/LiBSC18} proposed methods using meta-data, such as authorship information, titles, abstracts, keyword lists, and publication years. However, when applying such methods to real-world paper-writing tasks, there is a lack of consideration for the local context of a citation within a draft, which can potentially lead to suboptimal results. Context-based recommendations adopt a more practical concept that generates potential citations for an input context \cite{He:2010:CCR:1772690.1772734,He:2011:CRW:1935826.1935926}. Based on the context-based methodology, the HyperDoc2Vec \cite{DBLP:conf/acl/ShiSZZH18} proposed an embedding framework which further considers embedding with information of citation link between the local context in a citing paper and the content in a cited paper. Our previous study \cite{9123859} adapted the structural contexts in addition to citation link to further improve the recommendation performances. Context-based approaches could be potentially applicable in the real-world paper-writing process. 

However, the studies mentioned above still fail to take into consideration a number of essential characteristics of academic papers, which limits their usefulness.

\begin{enumerate}
    \item Scientific papers tend to follow the established ``IMRaD'' format (introduction, methods, results and discussion, and conclusions) \cite{10.1117/1.JMM.13.4.040101}, where each section of an article has a specific purpose. For example, the introduction section defines the topic of the paper in a broader context, the method section includes information on how the results were produced, and the results and discussion section presents the results. Therefore, citations used in each section should comply with the specific purpose of that section. For example, citations in the introduction section should support the main concepts of the paper, citations in the methods section should provide technical details, and citations in the results and discussion section should aim to compare results to those of other works. Therefore, recommendations of suitable citations for a given context should also consider the purpose of the corresponding section.
    
    \item Certain words and cited articles in a paper are much more closely related than other words and articles in the same paper. Capturing these interactions is essential for understanding a paper. For example, in Figure \ref{fig:concept}, the word ``recommendation'' is closely related to the words ``context-based,'' ``citations,'' and ``context,'' but has a weak relationship with the words ``adopt,'' ``more,'' and ``input.'' Additionally, a given word may have strong relatedness with some citations that appear in the paper. For example, the word ``recommendation'' has strong relatedness to citations ``\cite{DBLP:journals/access/LiBSC18}” and ``\cite{DBLP:conf/acl/ShiSZZH18}'' because both of these citations focus on recommendation algorithms. 
    
    \item Not every word or cited article has the same importance within a given paper. Important words and cited articles are more informative with respect to the topic of a paper. For example, in Figure \ref{fig:concept}, the words ``context-based,'' ``recommendations,'' ``citations,'' and ``context'' are more informative than the words ``adopt,'' ``more,'' or ``generates.'' The citation, ``\cite{DBLP:conf/acl/ShiSZZH18},'' may be more essential than ``\cite{DBLP:conf/ecir/JiaS18}'' because the former is related to context-based recommendations, while the latter is related to a different approach. 
 
\end{enumerate}

Adequate recommendations of citations for a manuscript should capture the relatedness and importance of words and cited articles in the context which needs citations, as well as the purpose of the section on which the writer is currently working. To this end, we propose a novel embedding-based neural network called dual attention model for citation recommendation (DACR) that is designed to capture the relatedness and importance of words in the context which needs citations and structural contexts in the manuscript, as well as the section for which the user is working. The core of the proposed neural network is composed of two attention mechanisms, namely self-attention and additive attention. The former captures the relatedness between contextual words and structural contexts, and the latter learns the importance of contextual words and structural contexts. Additionally, the proposed model embeds sections into an embedding space and utilizes the embedded sections as additional features for recommendation tasks.

\section{Related Work}

\subsection{Document Embedding}
Document embedding refers to the representation of words and documents as continuous vectors. Word2Vec \cite{DBLP:journals/corr/abs-1301-3781} was proposed as a shallow neural network for learning word vectors from texts while preserving word similarities. Doc2Vec \cite{DBLP:conf/icml/LeM14} is an extension of Word2Vec for embedding documents with content words. However, these two methods generally treat documents as ``plain texts,'' meaning that when they are applied to scholarly articles. This can lead to some essential information being lost (for example, citations and metadata in scientific papers), which in turn can lead to suboptimal recommendation results. Some more recent studies have attempted to remedy this issue. HyperDoc2Vec \cite{DBLP:conf/acl/ShiSZZH18} is a fine-tuning model for embedding additional citation relations. DocCit2Vec \cite{9123859} proposed by our previous work considers both structural contexts and citation relations. Regardless, some vital information is still not considered, such as the semantic of section headers and the relatedness and importance of word in the context requiring support of citations, which are included in this study.

\subsection{Citation Recommendation}
Citation recommendation refers to the task of finding relevant documents based on an input query. The query could be a collection of seed papers \cite{DBLP:conf/cscw/McNeeACGLRKR02,DBLP:conf/webi/GoriP06,DBLP:conf/jcdl/CarageaSMG13,DBLP:conf/asunam/KucuktuncSKC13,DBLP:conf/asunam/JiaS17}, and the recommendations are then generated via collaborative filtering \cite{DBLP:conf/cscw/McNeeACGLRKR02,DBLP:conf/jcdl/CarageaSMG13} or PageRank-based methods \cite{DBLP:conf/webi/GoriP06,DBLP:conf/asunam/KucuktuncSKC13,DBLP:conf/asunam/JiaS17}. Some studies \cite{DBLP:conf/lwa/AlzoghbiA0L15,DBLP:journals/access/LiBSC18} have proposed using meta-data, such as titles, abstracts, keyword lists, and publication years as query information. However, in real-world applications, when providing support for writing manuscripts, these techniques lack practicability. Context-based methods \cite{He:2010:CCR:1772690.1772734,He:2011:CRW:1935826.1935926,DBLP:conf/acl/ShiSZZH18,9123859} use a passage requiring support as a query to determine the most relevant papers, which can potentially enhance the paper-writing process. However, such methods may suffer from information loss because they do not consider sections within papers or the relative importance and relatedness of local context words.

\subsection{Attention Mechanisms}
Attention mechanism is commonly applied in the field of computer vision \cite{DBLP:conf/nips/TangSS14} and detects important parts of an image to improve prediction accuracy. This mechanism has also been adopted in the recent researches in text mining. For example, \cite{DBLP:conf/emnlp/LingTAFDBTL15} extended Word2Vec with a simple attention mechanism to improve word classification performance. Google’s BERT algorithm \cite{DBLP:conf/naacl/DevlinCLT19} uses multi-head attention and provides excellent performance for several natural language processing tasks. The method introduced in \cite{DBLP:conf/emnlp/WuWGQHX19} uses self-attention and additive attention to improve recommendation accuracy for news sources. 

\begin{figure}[tb]
  \centering
  \includegraphics[width=\textwidth]{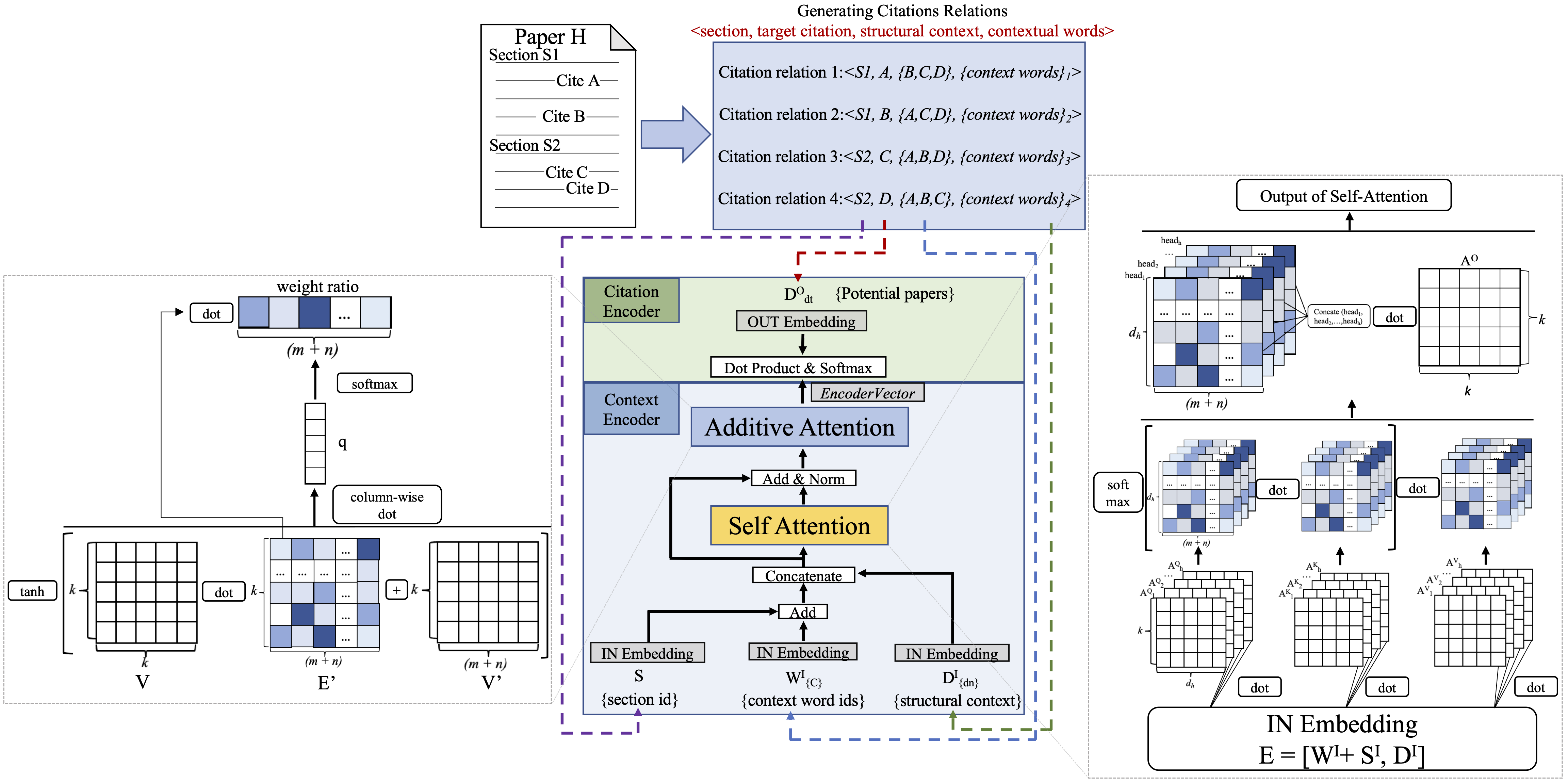}
  \caption{Architecture of DACR}
  \label{fig:dacr}
\end{figure}

\section{Preliminary} \label{sec:notation}
\subsection{Notations and Definitions} \label{sec:definition}
Academic papers can be treated as a type of hyper-document, in which citations are equivalent to hyperlinks. Based on paper modeling with citations \cite{DBLP:conf/acl/ShiSZZH18} and modeling of citations with structural contexts \cite{9123859}, we introduce a novel modeling with citations, structural contexts, and sections.

\begin{theorem}[Academic Paper]
Let $\mathit{w} \in \mathit{W}$ represent a word from a vocabulary, $\mathit{W}$, where $\mathit{s} \in \mathit{S}$ represents a section from a section header collection, $\mathit{S}$, and $\mathit{d}\in\mathit{D}$ represents a document ID (paper DOI) from an ID collection, $\mathit{D}$. The textual information of a paper, $\mathit{H}$, is represented as a sequence of words, sections, and IDs of cited documents (\textit{i.e.}, $\mathit{\hat{W}} \cup \mathit{\hat{S}} \cup \mathit{\hat{D}}$, where $\mathit{\hat{W}} \subseteq \mathit{W}$, $\mathit{\hat{S}} \subseteq \mathit{S}$, and $ {\hat{D}} \subseteq \mathit{D}$). 
\end{theorem}

\begin{theorem}[Citation Relationships]\label{def2}
The citation relationships, $\mathcal{C}$, (see Figure \ref{fig:dacr}) in a paper, $\mathit{H}$, are expressed by a tuple, $\langle \mathit{s}, \mathit{d_t},  \mathit{D_n}, \mathit{C} \rangle$, where $d_t \in \mathit{\hat{D}}$ represents a target citation, $\mathit{\hat{D}}$ represents the id of all the cited documents from $\mathit{H}$, $\mathit{C} \subseteq \mathit{\hat{W}}$ is the local context surrounding $\mathit{d_t}$, and $\mathit{s} \in \mathit{\hat{S}}$ is the title of the section in which the contextual words appear. If other citations exist within the same manuscript, then they are defined as \textit{``structural contexts''} and denoted by $\mathit{D_n}$, where $\{\mathit{d_n} | \mathit{d_n} \in \mathit{\hat{D}}, \mathit{d_n} \neq \mathit{d_t}\}$.
\end{theorem}

\subsection{Problem Definition}
Embedding matrices are denoted as $\mathbf{D} \in \mathbb{R}^{k \times |D|}$ for documents, $\mathbf{W} \in \mathbb{R}^{k \times |W|}$ for words, and $\mathbf{S} \in \mathbb{R}^{k \times |S|}$ for sections. The \textit{i}-th column of $\mathbf{D}$, denoted by $\mathbf{d_i}$, is a $\mathit{k}$-dimensional vector representing document $\mathit{d_i}$. Additionally, the \textit{j}-th column of $\mathbf{W}$ is a \textit{k}-dimensional vector for word $\mathit{w_j}$, and the \textit{s}-th column of $\mathbf{S}$ is a \textit{k}-dimensional vector for section $\mathit{s}$.

The proposed model initializes two embedding matrices (IN and OUT) for documents (i.e., $\mathbf{D^I}$ and $\mathbf{D^O}$), a word embedding matrix, $\mathbf{W^I}$, and a section embedding matrix, $\mathbf{S^I}$. A column vector from $\mathbf{D^I}$ represents the role of a document as a structural context and a column vector from $\mathbf{D^O}$ represents the role of a document as a citation (the implementation details of the experiment in Section \ref{sec:results} explain this in more detail). The word embedding matrix, $\mathbf{W^I}$, and section embedding matrix, $\mathbf{S^I}$, are initialized for all words of the word vocabulary and all sections of the section header collection.

The goal of this model is to optimize the following objective function:
\begin{equation}
    \max_{\mathbf{D^{I}}, \mathbf{D^{O}}, \mathbf{W^{I}},\mathbf{S^{I}}} 
    \frac{1}{|\mathcal{C}|} \sum_{\langle \mathit{s}, \mathit{d_t}, \mathit{D_n}, \mathit{C} \rangle \in \mathcal{C}}
    \log P(\mathit{d_t| \mathit{s},\mathit{D_n}, \mathit{C} }).
\end{equation}

\section{Dual Attention Model for Citation Recommendation}
An overview of the proposed DACR approach is presented in Figure \ref{fig:dacr}. DACR is composed of two main components: a context encoder (Section \ref{sec:context_encoder}) for encoding contextual words, sections, and structural contexts into a fixed-length vector and a citation encoder (Section \ref{sec:citation_encoder}) for predicting the probability of a target citation.

\subsection{Context Encoder} \label{sec:context_encoder}
The context encoder takes three inputs, namely, context words, sections, and structural contexts, from citation relationships. The encoder contains three layers: an embedding layer for converting words and documents (structural contexts) into vectors, a self-attention layer with an Add\&Norm sub-layer \cite{DBLP:conf/nips/VaswaniSPUJGKP17} for capturing the relatedness between words and structural contexts, and an additive attention layer \cite{DBLP:conf/emnlp/WuWGQHX19} for recognizing the importance of each word and structural context.
\subsubsection{IN Embedding, Add and Concatenation layer}
The IN embedding layer initially generates three embedding matrices $\mathbf{D^I}$, $\mathbf{W^I}$, and $\mathbf{S^I}$ for the document collection, word vocabulary and the section header collection. For a given citation relationship, the one-hot vectors of structural contexts, context words, and sections are projected with the three embedding matrices, denoted as $\mathbf{D^I}_{\{D_n\}}$, $\mathbf{W^I}_{\{C\}}$, and $\mathbf{S^I}_s$. The projected section vectors are then added to the word vectors (each word vector is added to a section vector), and the resultant matrix is denoted as $\mathbf{W'}$. $\mathbf{W'}$ and $\mathbf{D^I}_{\{D_n\}}$ are then concatenated column-wise and form one matrix, i.e., $[\mathbf{w'_1},...,\mathbf{w'_m}, \mathbf{d^I_1},...,\mathbf{d^I_n}]$, and denoted as $\mathbf{E}$, where $m$ is the number of input context words and $n$ is the number of input structural contexts.


\subsubsection{Self-attention Mechanism with Add\&Norm}
Self-attention \cite{DBLP:conf/nips/VaswaniSPUJGKP17} is utilized to capture the relatedness between input context words and structural contexts. It applies scaled dot-product attention in parallel for a number of heads, to allow the model to jointly consider interactions from different representation sub-spaces at different positions.

The $k$-dimensional embedding matrix, $\mathbf{E}$, from the last layer is first transposed and projected with three linear projections ($\mathbf{A}_i^Q$,$\mathbf{A}_i^K$, and $\mathbf{A}_i^V$) to a $d_h$ dimensional space, where $d_h = k / h$, $i \in \{1...h\}$, and $h$ denotes the number of heads. The $\mathbf{E}$ matrix is projected $h$ times, and each projection is called a ``head''. At each projection (i.e., within a ``head''), the dot products of the first two projected versions of $\mathbf{E}$ with $\mathbf{A}_i^Q$ and $\mathbf{A}_i^K$ are computed, and divided by $\sqrt{d_h}$. Subsequently, softmax is applied to obtain the resulting weight matrix with dimensions of $(m+n) * (m+n)$, i.e., $softmax(\frac{\mathbf{E}^{T} \mathbf{A}_i^Q \cdot (\mathbf{E}^{T} \mathbf{A}_i^K)^T}{\sqrt{d_h}})$ , where $(m+n)$ is the total number of input context words and structural contexts. This weight matrix represents the relatedness between the input words and articles. The dot product of the weight matrix and the third projected version of $\mathbf{E}$, i.e., $\mathbf{E}^T \mathbf{A}_i^V$, is computed as the output matrix of the head, denoted as $\mathbf{head}_i$. The $h$ numbers of the output head matrices are concatenated column-wise and projected again with $\mathbf{A}^O$ to yield the final output matrix. The computation procedure is represented as follows:
\begin{equation}
    SelfAttention(\mathbf{E})=Concat(\mathbf{head}_1,...,\mathbf{head}_h)\mathbf{A}^O,
\end{equation}
\begin{equation}
    \mathbf{head}_i=softmax(\frac{\mathbf{E}^{T} \mathbf{A}_i^Q \cdot (\mathbf{E}^{T} \mathbf{A}_i^K)^T}{\sqrt{d_h}})\cdot(\mathbf{E}^{T} \mathbf{A}_i^V),
\end{equation}
where $\mathbf{A}^O \in \mathbb{R}^{k \times k}$, $\mathbf{A}_i^Q \in \mathbb{R}^{k \times d_h}$, $\mathbf{A}_i^K \in \mathbb{R}^{k \times d_h}$, and $\mathbf{A}_i^V \in \mathbb{R}^{k \times d_h}$ are projection parameters. $d_h$ is the embedding dimension of the heads, $h$ is the number of heads, and $k=d_h \times h$, where $k$ is the dimension of the embedding vectors. The output matrix of the self-attention mechanism is then transposed and added to the original $\mathbf{E}$ matrix. Next, dropout is applied \cite{DBLP:journals/corr/abs-1207-0580} to avoid over-fitting, and applied with layer normalization \cite{DBLP:journals/corr/BaKH16} to facilitate the convergence of the model during training. The final output matrix is denoted as $\mathbf{E'}$.

\subsubsection{Additive Attention Mechanism}
The additive attention layer \cite{DBLP:conf/emnlp/WuWGQHX19} is utilized to recognize informative contextual words and structural contexts. It takes matrix $\mathbf{E'}$ from the last layer as input, whereby each column represents the vector of a word or document. The weight of each item is computed as follows:

\begin{equation} \label{eqn:weights}
    \mathbf{Weight} = \mathbf{q}^T \cdot \tanh{(\mathbf{V} \cdot \mathbf{E}'  + \mathbf{V}')},
\end{equation}

where $\mathbf{V} \in \mathbb{R}^{k \times k}$ is the projection parameter matrix, $\mathbf{V}' \in \mathbb{R}^{k \times (n+m)}$ is the bias matrix, and $\mathbf{q}$ ($k$-dimensional) is a parameter vector. The $\mathbf{Weight}$ vector is a row vector of dimension $(m+n)$, where each column represents the weight of a corresponding word or document. The $\mathbf{Weight}$ vector is applied with the dropout technique to avoid over-fitting.

The output, $\mathbf{EncoderVector}$, is the dot product of the softmaxed $\mathbf{Weight}$ vector and input matrix, $\mathbf{E'}$, where all rows of the embedding vectors are weighted and summed, as illustrated below:
\begin{equation} \label{eqn:encodervec}
    \mathbf{EncoderVector} = \mathbf{E}' \cdot softmax(\mathbf{Weight}^{T}).
\end{equation}

\subsection{Citation Encoder} \label{sec:citation_encoder}

The citation encoder is designed to predict potential citations by calculating the probability score between an OUT document matrix, $\mathbf{D^O}$, and the $\mathbf{Encoder Vector}$ from the context encoder, which is defined as follows:
\begin{equation}
    \hat{\mathbf{y}}=\mathbf{Encoder Vector}^T \cdot \mathbf{D^O}.
\end{equation}
The scores are then normalized using the softmax function as follows:
\begin{equation}
    \mathbf{p}=softmax(\hat{\mathbf{y}}).
\end{equation}

\subsection{Model Training and Optimization}

We adopted a negative sampling training strategy \cite{DBLP:conf/nips/MikolovSCCD13} to speed up the training process for DACR. In each iteration, it generates a positive sample (correctly cited paper) and $n$ negative samples. Therefore, the calculated probability vector, $\mathbf{p}$, is composed of $[p_{positive}, p_{negative-1},p_{negative-2},...,p_{negative-n}]$. The loss function computes the negative log-likelihood of the probability of a positive sample, as follows:
\begin{equation}
    \mathcal{L}= - \log(p_{positive}) + \sum^n_{i=1} \log({p_{negative-i}}).
\end{equation}

Stochastic gradient descent (SGD) \cite{DBLP:conf/icml/SutskeverMDH13} is used to optimize the model.

\section{Experiments}
We evaluated the recommendation performance of our model and five baseline models on two datasets, namely DBLP and ACL Anthology \cite{DBLP:conf/acl/ShiSZZH18}. The recall, mean average precision (MAP), mean reciprocal rank (MRR), and normalized discounted cumulative gain (nDCG) are reported for a comparison of the models. These values are summarized in Table \ref{tab:result}. Additionally, we proved the effectiveness of adding information about sections, relatedness, and importance, as shown in Figure \ref{fig:effect}.

\subsection{Dataset Overview} \label{sec:data}
The larger dataset, DBLP \cite{DBLP:conf/acl/ShiSZZH18}, contains 649,114 full paper texts with 2,874,303 citations (approximately five citations per paper) in the field of computer science. The ACL Anthology dataset \cite{DBLP:conf/acl/ShiSZZH18} is smaller, containing 20,408 texts with 108,729 citations; however, it has a similar number of citations per paper (about five per paper) to the DBLP dataset. We split the datasets into a training dataset, for training the document, word, and section vectors, and test dataset with papers containing more than one citation published in the last few years for recommendation experiments. An experimental overview is provided in Table \ref{table:data}.

\subsection{Document Preprocessing}
The texts were pre-processed using ParsCit \cite{DBLP:conf/lrec/CouncillGK08} to recognize citations and sections. In-text citations were replaced with the corresponding unique document IDs in the dataset. Section headers often have diverse names. For example, many authors name the ``methodology'' section using customized algorithm names. Therefore, we replaced all section headers with fixed generic section headers using ParsLabel \cite{DBLP:journals/ijdls/LuongNK10}. Generic headers from ParsLabel are \textit{abstract, background, introduction, method, evaluation, discussions, and conclusions}. If ParsLabel is not able to recognize a section, we label it as ``\textit{unknown}.'' Detailed information for each section is listed in Table \ref{table:data}.

\subsection{Implementation and Settings} \label{sec:setting}
DACR was developed using PyTorch 1.2.0 \cite{DBLP:conf/nips/PaszkeGMLBCKLGA19}. In our experiments, word and document embeddings were pre-trained using DocCit2Vec with an embedding size of 100, a window size of 50 (also known as the length of the local context, i.e. 50 words before and after a citation), a negative sampling value of 1000, and 100 iterations (default settings in \cite{9123859}). The word vectors for generic headers, such as ``introduction'' and ``method,'' were selected as pre-trained vectors for the section headers. DACR was implemented with 5 heads, 100 dimensions for the query vector, and a negative sampling value of 1000. The SGD optimizer was implemented with a learning rate of 0.001, a batch size of 100, and 100 iterations for the DBLP dataset, or 300 iterations for the ACL Anthology dataset. To avoid over-fitting, we applied a 20\% dropout in the two attention layers. 

Word2Vec and Doc2Vec were implemented using Gensim 2.3.0 \cite{rehurek_lrec}, and HyperDoc2Vec and DocCit2Vec were developed based on Gensim. All baseline models were initialized with an embedding size of 100, a window size of 50, and default values for the remaining parameters.
\begin{table}[tb]
\caption{Statistics of the datasets}
\label{table:data}
\resizebox{\textwidth}{!}{%
\begin{tabular}{lllll|lllllllll}
\hline
\multicolumn{5}{c|}{\textbf{Overview of the dataset}}                                                      & \multicolumn{9}{c}{\textbf{Number of sections in the dataset}}                                                                                                                                      \\ \hline
                               &                           & \textbf{All} & \textbf{Train} & \textbf{Test} & \textbf{Generic Section} & \textbf{Abstract} & \textbf{Background} & \textbf{Introduction} & \textbf{Method} & \textbf{Evaluation} & \textbf{Discussion} & \textbf{Conclusions} & \textbf{Unknown} \\
\multirow{2}{*}{\textbf{DBLP}} & \textbf{No. of Docs}      & 649,114      & 630,909        & 18,205        & \textbf{Train}           & 617,402           & 9,589               & 452,430               & 3,226,521       & 153,737             & 19,738               & 435,514              & 155,777          \\
                               & \textbf{No. of Citations} & 2,874,303    & 2,770,712      & 103,591       & \textbf{Test}            & 5,243             & 155                 & 6,437                 & 25,956          & 1,312               & 200                  & 1875                 & 58,975           \\ \hline
\multirow{2}{*}{\textbf{ACL}}  & \textbf{No. of Docs}      & 20,408       & 14,654         & 1,563         & \textbf{Train}           & 11,725            & 114                 & 9,973                 & 42,749          & 4,186               & 442                  & 9,456                & 847              \\
                               & \textbf{No. of Citations} & 108,729      & 79,932         & 28,797        & \textbf{Test}            & 3,789             & 33                  & 3,429                 & 12,625          & 1,587               & 159                  & 3,186                & 0                \\ \hline
\end{tabular}
}
\end{table}

\begin{table}[!t]
\renewcommand{\thefootnote}{\fnsymbol{footnote}}
\centering
\caption{Citation recommendation results (** statistically significant at 0.01 significance level)} 
\label{tab:result}
\resizebox{\textwidth}{!}{%
\begin{tabular}{l|llll|llll}
\hline
\multicolumn{1}{c|}{\multirow{2}{*}{\textbf{Model}}} & \multicolumn{4}{c|}{\textbf{DBLP}}                                        & \multicolumn{4}{c}{\textbf{ACL}}                                          \\ \cline{2-9} 
\multicolumn{1}{c|}{}                                & \textbf{Recall@10} & \textbf{MAP@10} & \textbf{MRR@10} & \textbf{nDCG@10} & \textbf{Recall@10} & \textbf{MAP@10} & \textbf{MRR@10} & \textbf{nDCG@10} \\ \hline
W2V (case 1)                                         & 20.47              & 10.54           & 10.54           & 14.71            & 27.25              & 13.74           & 13.74           & 19.51            \\
W2V (case 2)                                         & 20.46              & 10.55           & 10.55           & 14.71            & 26.54              & 13.55           & 13.55           & 19.19            \\
W2V (case 3)                                         & 20.15              & 10.40           & 10.40           & 14.49            & 26.06              & 13.21           & 13.21           & 18.66            \\
D2V-nc (case 1)                                      & 7.90               & 3.17            & 3.17            & 4.96             & 19.92              & 9.06            & 9.06            & 13.39            \\
D2V-nc (case 2)                                      & 7.90               & 3.17            & 3.17            & 4.96             & 19.89              & 9.06            & 9.06            & 13.38            \\
D2V-nc (case 3)                                      & 7.91               & 3.17            & 3.17            & 4.97             & 19.89              & 9.07            & 9.07            & 13.38            \\
D2V-cac (case 1)                                     & 7.91               & 3.17            & 3.17            & 4.97             & 20.51              & 9.24            & 9.24            & 13.68            \\
D2V-cac (case 2)                                     & 7.90               & 3.17            & 3.17            & 4.97             & 20.29              & 9.17            & 9.17            & 13.58            \\
D2V-cac (case 3)                                     & 7.89               & 3.17            & 3.17            & 4.97             & 20.51              & 9.24            & 9.24            & 13.69            \\
HD2V (case 1)                                        & 28.41              & 14.20           & 14.20           & 20.37            & 37.53              & 19.64           & 19.64           & 27.20            \\
HD2V (case 2)                                        & 28.42              & 14.20           & 14.20           & 20.38            & 36.83              & 19.62           & 19.62           & 27.18            \\
HD2V (case 3)                                        & 28.41              & 14.20           & 14.20           & 20.37            & 36.24              & 19.32           & 19.32           & 26.79            \\
DC2V (case 1)                                        & 44.23              & 21.80           & 21.80           & 31.34            & 36.89              & 20.44           & 20.44           & 27.72            \\
DC2V (case 2)                                        & 40.31              & 20.16           & 20.16           & 28.69            & 33.71              & 18.47           & 18.47           & 25.17            \\
DC2V (case 3)                                        & 40.37              & 19.02           & 19.02           & 26.84            & 31.14              & 16.97           & 16.97           & 23.20            \\ \hline
DACR (case 1)                                        & \textbf{48.96}\footnotemark[7]     & \textbf{23.25}\footnotemark[7]  & \textbf{23.25}\footnotemark[7]  & \textbf{33.93}\footnotemark[7]  & \textbf{42.43}\footnotemark[7]      & \textbf{22.92}\footnotemark[7]   & \textbf{22.92}\footnotemark[7]   & \textbf{31.64}\footnotemark[7]    \\
DACR (case 2)                                        & \textbf{45.39}\footnotemark[7]     & \textbf{22.32}\footnotemark[7]  & \textbf{22.32}\footnotemark[7] & \textbf{31.98}\footnotemark[7]  & \textbf{40.13}\footnotemark[7]      & \textbf{21.93}\footnotemark[7]   & \textbf{21.93}\footnotemark[7]   & \textbf{30.04}\footnotemark[7]    \\
DACR (case 3)                                        & \textbf{42.32}\footnotemark[7]     & \textbf{21.39}\footnotemark[7]  & \textbf{21.39}\footnotemark[7]  & \textbf{30.22}\footnotemark[7]   & \textbf{38.01}\footnotemark[7]      & \textbf{20.84}\footnotemark[7]   & \textbf{20.84}\footnotemark[7]   & \textbf{28.45}\footnotemark[7]    \\ \hline
\end{tabular}
}
\end{table}

\subsection{Recommendation Evaluation} \label{sec:results}
We designed three usage cases to simulate real-world scenarios:
\begin{itemize}
    \item Case 1: In this case, we assumed the manuscript was approaching its completion phase, meaning the writer had already inserted the majority of their citations into the manuscript. Based on the leave-one-out approach, the task was to predict a target citation, by providing the contextual words (50 words before and after the target citation), structural contexts (the other cited papers in the source paper), and section header as input information for DACR.
    
    \item Case 2: Here, we assumed that some existing citations were invalid because they were not available in the dataset, i.e., the author had made typographical errors or the manuscript was in an early stage of development. In this case, given a target citation, its local context and section header, we randomly selected structural contexts to predict a target citation. The random selection was implemented using the build-in Python3 \textit{random} function. All case 2 experiments were conducted three times to determine the average results to rule out biases.
    
    \item Case 3: It is assumed that the manuscript was in an early phase of development, where the writer has not inserted any citations or all existing citations are invalid. Only context words and section headers were utilized for the prediction of the target citation (no structural contexts were used).
\end{itemize}

To conduct recommendation via DACR, an encoder vector was initially inferred using the trained model with inputs of cases 1, 2, and 3, and then, the OUT document vectors were ranked based on dot products.

Five baseline models were adapted for comparison with DACR. As the baseline models do not explicitly consider section information, information on the section headers were neglected in the inputs.
\begin{enumerate}
    \item \textbf{Citations as words via Word2Vec (W2V)} This method was presented in \cite{DBLP:journals/tvcg/BergerMS17}, where all citations were treated as special words. The recommendation of documents was defined as ranking OUT word vectors of documents relative to the averaged IN vectors of context words, and structural contexts via dot products. The word vectors were trained using the Word2Vec CBOW algorithm.
    \item \textbf{Citations as words via Doc2Vec (D2V-nc)}\cite{DBLP:journals/tvcg/BergerMS17}. The citations were removed in this method, and the recommendations were made by ranking the IN document vectors via cosine similarity relative to the vector inferred from the learnt model by taking context words and structural contexts as input (this method results in better performance than the dot product). The word and document vectors were trained using Doc2Vec PV-DM.
    \item \textbf{Citations as content via Doc2Vec (D2V-cac)} \cite{DBLP:conf/acl/ShiSZZH18}. In this method, all context words around a citation were copied into the cited document as supplemental information. The recommendations were made based on cosine similarity between the IN document vectors and inferred vector from the learnt model. The vectors were trained using Doc2Vec PV-DM.
    \item \textbf{Citations as links via HyperDoc2Vec (HD2V)} \cite{DBLP:conf/acl/ShiSZZH18}. In this method, citations were treated as links pointing to target documents. The recommendations were made by ranking OUT document vectors relative to the averaged IN vectors of input contextual words based on dot products. The embedding vectors were pre-trained by Doc2Vec PV-DM using default settings.
    \item \textbf{Citations as links with structural contexts via DocCit2Vec (DC2V)} \cite{9123859}. The recommendations were made by ranking OUT document vectors relative to the averaged IN vectors of input contextual words and structural contexts based on dot products. The embedding vectors were pre-trained by Doc2Vec PV-DM with default settings.
\end{enumerate}

There are three main conclusions that can be drawn from Table \ref{tab:result}. First, DACR outperforms all baseline models at 1\% significance level across all evaluation scores for all cases and datasets. This implies that the additionally included combined information: namely sections, relatedness, and importance, are essential for predicting useful citations. The effectiveness of each added information is presented in Section \ref{sec:effectvess}.

Second, performance increases when additional information is preserved in the embedding vectors. When comparing Word2Vec, HyperDoc2Vec, DocCit2Vec, and DACR, Word2Vec only preserves contextual information, HyperDoc2Vec considers citations as links, DocCit2Vec includes structural contexts, and DACR exploits the internal structure of a scientific paper to extract richer information. The evaluation scores increase with the amount of information preserved, indicating that overcoming information loss in embedding algorithms is helpful for recommendation tasks.

Third, DACR is effective for both the large (DBLP) and medium (ACL Anthology) sized datasets. However, we also realized that the smaller dataset requires higher iterations for the model to produce effective results. It is presumed that more iterations of training can compensate for a lack of diversity in the training data.


The performance of DACR could be further improved by more accurately recognizing section headers. Moreover, we determined that some labels were incorrectly recognized or unable to be recognized by ParsLabel. Therefore, we will work on improving the accuracy of section recognition in future work.

\subsection{Effectiveness of Adding Sections, Relatedness, and Importance}
\label{sec:effectvess}
\begin{figure}[tb]
  \centering
  \includegraphics[width=0.6\columnwidth]{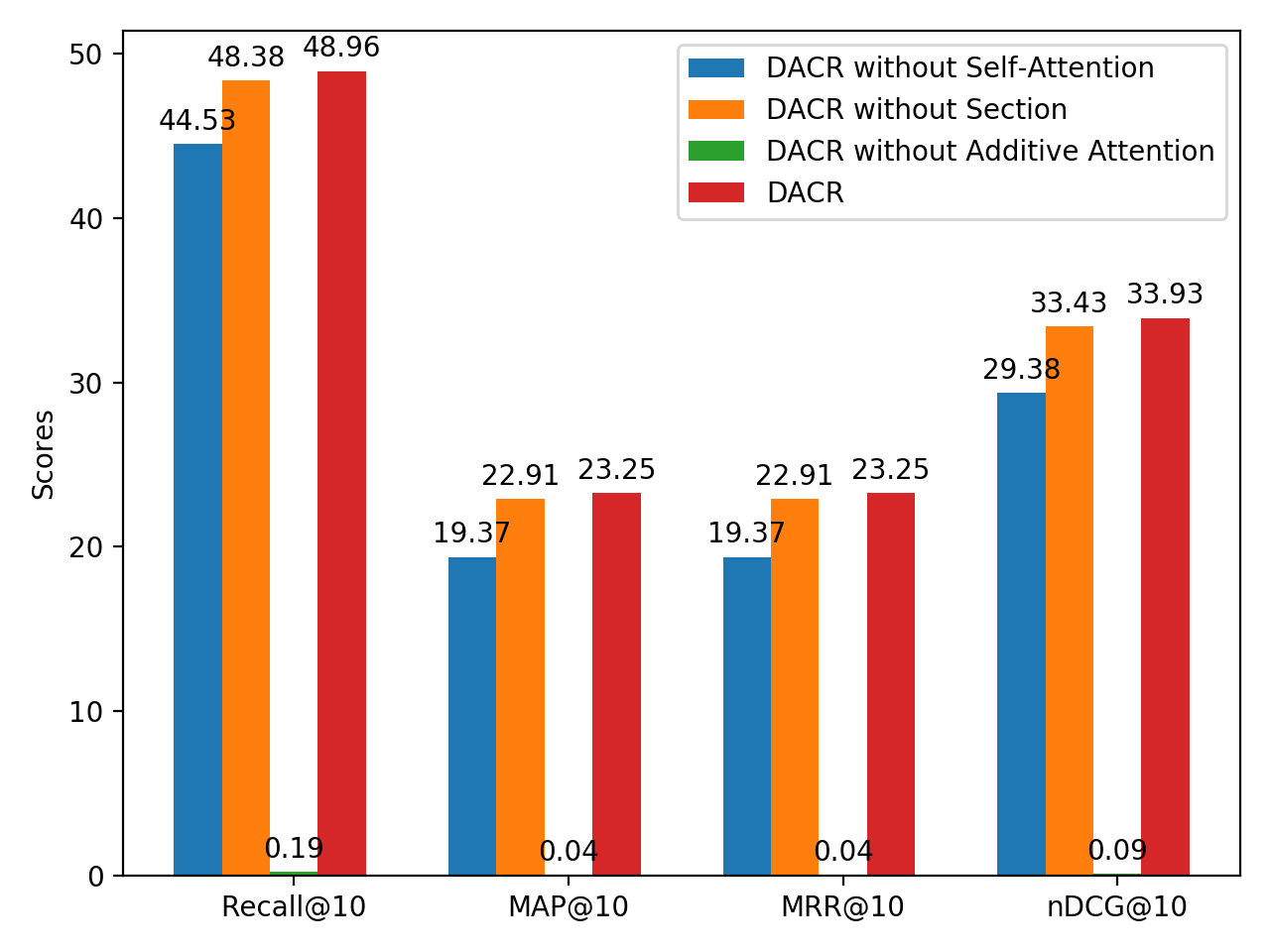}
  \caption{Effectiveness of adding sections, relatedness, and importance}
  \label{fig:effect}
\end{figure}
In this section, we explore the effectiveness of adding the following information: section headers, relatedness, and importance. We run three modified DACR models without the corresponding layer, for example, removing the section embedding layer for verifying the effectiveness of section information, removing the self-attention layer for determining the relatedness between contextual words and articles, and removing additive attention for demonstrating the importance of context. We present the scores of recall, MAP, MRR, and nDCG at 10 for case 1 on the DBLP dataset for comparison, which are illustrated in Figure \ref{fig:effect}.

Both models, DACR without self-attention and DACR without additive attention perform significantly worse than the full model of DACR, whereas the performance of DACR without section information drops by a minor margin. Three conclusions can be drawn from these facts.


Firstly, all modified models performed poorer than the full model, which supports our hypothesis: sections, relatedness, and importance between contextual words and articles are important for recommending useful citations. The relatedness information is more beneficial than section information, which is evident when comparing DACR without section and DACR without self-attention.

Secondly, the primary reason for the 0-close scores of the model without additive attention is that the losses of the model did not converge without the additive attention layer. Therefore, we consider that the additive attention has a two-fold purpose: ensuring convergence and learning the importance of context.

Lastly, only appropriate combinations of information and neural network layers lead to optimal solutions, as deficits in any of the three types of information (section, relatedness, and importance, or attention layers) result in low performance.

\section{Conclusions and Future Work}
In this study, we proposed a citation recommendation model with dual attention mechanisms. This model aims to simplify real-world paper-writing tasks by alleviating the issue of information loss in existing methods. Our model considers three types of essential information: section for which a user is working and need to insert citations, relatedness between the local context words and structural contexts, and their importance. The core of the proposed model is composed of two attention mechanisms: self-attention for capturing relatedness and additive attention for learning importance. Extensive experiments demonstrated the effectiveness of the proposed model in designed scenarios intended to mimic the real world scenarios as well as the efficiency of the proposed neural network.

In future work, we will first attempt to improve the accuracy of recognizing section headers to improve the usability and performance of the algorithm. Second, we will include additional paper-related information in the model, such as word positions. Third, we will explore more sophisticated neural network architectures to improve accuracy and reduce the training time of the model.

\section*{Acknowledgements}
This research has been supported in part by JSPS KAKENSHI under Grant Number 19H04116 and by MIC SCOPE under Grant Numbers 201607008 and 172307001.

\bibliographystyle{coling}
\bibliography{section_model}

\end{document}